\documentclass[aps,prb,reprint,superscriptaddress,showpacs]{revtex4-1}

\usepackage[english]{babel}
\usepackage{ragged2e}
\justifying
\usepackage{amsmath}
\usepackage{tikz}
\usetikzlibrary{shapes,arrows}
\usepackage{verbatim}
\usepackage{graphicx}
\usepackage{float}
\usepackage[caption=false]{subfig}
\usepackage{amsfonts}
\usepackage{amssymb}
\usepackage{graphicx}
\usepackage{hyperref}
\begin{document}

\title{Exact two-body quantum dynamics of an electron-hole pair in semiconductor coupled quantum wells: a time-dependent approach}

\author{Federico Grasselli}
\email[]{federico.grasselli@unimore.it}

\affiliation{Dipartimento di Scienze Fisiche, Informatiche e Matematiche, Universit\`a degli Studi di Modena e Reggio Emilia, Via Campi 213/a, Modena, Italy}
\affiliation{S3, CNR-Istituto Nanoscienze, Via Campi 213/a, Modena, Italy}

\author{Andrea Bertoni}
\email[]{andrea.bertoni@nano.cnr.it}

\affiliation{S3, CNR-Istituto Nanoscienze, Via Campi 213/a, Modena, Italy}

\author{Guido Goldoni}
\email[]{guido.goldoni@unimore.it}

\affiliation{Dipartimento di Scienze Fisiche, Informatiche e Matematiche, Universit\`a degli Studi di Modena e Reggio Emilia, Via Campi 213/a, Modena, Italy}
\affiliation{S3, CNR-Istituto Nanoscienze, Via Campi 213/a, Modena, Italy}

\begin{abstract}
We simulate the time-dependent coherent dynamics of a spatially indirect exciton -an electron-hole pair with the two particles confined in different layers- in a GaAs coupled quantum well system. We use a unitary wave-packet propagation method taking into account in full the four degrees of freedom of the two particles in a two-dimensional system, including both the long-range Coulomb attraction and arbitrary two-dimensional electrostatic potentials affecting the electron and/or the hole separately. The method has been implemented for massively parallel architectures to cope with the huge numerical problem, showing good scaling properties and allowing evolution for tens of picoseconds. We have investigated both transient time phenomena and asymptotic time transmission and reflection coefficients for potential profiles consisting of i) extended barriers and wells and ii) a single-slit geometry. We found clear signatures of the internal two-body dynamics, with transient phenomena in the picosecond time-scale which might be revealed by optical spectroscopy. Exact results have been compared with mean-field approaches which, neglecting dynamical correlations by construction, turn out to be inadequate to describe the electron-hole pair evolution in realistic experimental conditions.
\end{abstract}


\maketitle

\section{Introduction}

Electronic bilayer systems have exposed a huge amount of new physics driven by inter-layer Coulomb interactions. A few examples are fractional quantum Hall effect states in semiconductor\cite{Eisenstein1992,Suen1992} and graphene bilayers,\cite{Novoselov2006} phase transitions in Quantum Hall ferromagnets,\cite{Piazza1999} complex Wigner crystal ordering\cite{Goldoni1997}. Electron-hole bilayers gained importance in their own right. Excitons -bound electron-hole quasi-particles with a bosonic character- have long being predicted to undergo quantum condensation.\cite{Bryant_PRB93, Butov_NatureA02} Recently, signatures of condensation have been found in systems of spatially indirect excitons (IXs), electron-hole pairs optically excited in semiconductor coupled quantum well systems, with the two charges confined in different layers by a static electric field.\cite{Butov_JPCM04,High_NATURE12, Alloing_EPL14}

In a different perspective, IXs are at the heart of a new class of opto-electronic devices, made possible thanks to the small electron-hole overlap which extends their intrinsic lifetime from nanoseconds\cite{Colocci_EPL90} to microseconds.\cite{Butov_JPCM04,Gartner_APL06} Indeed, although IXs are neutral excitations, they carry a large finite electric dipole which can be used to drive the evolution of IXs in the coupled quantum well (CQW) planes by electric field gradients generated, e.g., by metallic gates. Finally, since a bias normal to the QW-plane can control the overlap of the pair, IX recombination can be induced at arbitrary time, thereby 'measuring' the result of the evolution. IX gases have been exploited to demonstrate several functionalities, such as fast data storage,\cite{Winbow_NL07,Winbow_JAP08} acceleration with electrostatic ramps\cite{Gartner_APL06,Leonard_APL12} and interdigital devices,\cite{Winbow_PRL11} field effect transistors.\cite{High_Science11072008} Furthermore, trapping of single IXs has been recently demonstrated,\cite{Schinner_PRL13} opening the way to single IX electronics. This requires the development of theoretical concepts to describe the evolution of IX wave-packets in complex electrostatic fields.

Scattering of a composite quantum objects with internal degrees of freedom\cite{SaitoJPCM94} (DoFs) like an IX in the presence of an electrostatic field gradient,\cite{SaitoPRB95} is an important topic in its own, with applications in molecular and nuclear physics (see, e.g., Refs.~\onlinecite{Bertulani2015,AhsanPRC10,KuperinJMP92,ShotterPRC11}, and references therein). Indeed, in the presence of a scattering potential, energy can be transferred between the center-of-mass (CM) kinetic energy and internal excitations, which may strongly influence the transmission and reflection probabilities. However, due to the difficulty to evolve the quantum equations of motion for several DoFs with open boundaries, exact calculations are often limited to idealized situations, such as collinear scattering, purely one-dimensional (1D) systems, and/or very simple potential profiles.\cite{SaitoJPCM94,AhsanPRC10,LugovskoyPRA13,HnybidaPRA08,PenkovJETP00,KavkaPRA10} Indeed, the numerical approach scales exponentially with the number of DoFs. For realistic situations, such as the  3D problem of colliding molecules with complex inter-molecular interactions, mean-field methods have been applied.\cite{BeckTDH2000}
IXs in CQWs are an important system from this point of view, since in principle their evolution could  be probed by accurate time-dependent optical means, not only at asymptotic times, but also during the scattering event. Recently, we used an idealized 1D model to study the evolution of IX wave-packets under the action of external fields, taking explicitly into account the internal structure of the electron-hole pair.\cite{GrasselliJCP15} Our model allowed to investigate different regimes/potential profiles where inter-particle Coulomb correlations may lead to internal excitations or even dissociation, as a result of scattering. However, such 1D calculations are too simplistic to be applied to realistic CQW quasi-two-dimensional (2D) systems, where electrons and holes evolve in a complex 3D structure.

In this paper we report time-dependent simulations of the coherent dynamics of a single IX wave-packet in a semiconductor CQW structure with complex in-plane electrostatic potentials, taking into account the 3D structure of realistic devices through a 2D+1D effective model. We used a unitary wave-packet propagation which includes the long-range electron-hole Coulomb interaction and arbitrary scattering potentials acting on the electron and the hole separately. The large numerical problem to simulate exactly the present four-DoF system has been tackled by the Fourier split-step method implemented on a massively parallel architecture. This allowed propagation for tens of picoseconds in typical potential landscapes. We have investigated potential profiles consisting of extended barriers/wells, and a single slit geometry, finding genuine signatures of the two-body dynamics in realistic experimental conditions, with transient phenomena in the picosecond time-scale.  Our results could be directly compared with time resolved optical spectroscopy. We have compared the unitary evolution results with a mean-field approach at different levels of approximation, the so-called rigid exciton (RIX) model and the time-dependent Hartree (TDH) method. The comparison shows that a mean-field approach is in general inadequate to describe the electron-hole pair evolution in realistic samples, thereby showing that IX dynamics in CQW might be a particularly interesting system to investigate correlation effects and to test theoretical modeling.

In Sec.~\ref{sec:theory} we define our Hamiltonian description of an IX in a typical semiconductor CQW (\ref{sec:model}) and we provide details on the full (\ref{sec:fullpropagation}) and mean-field (\ref{sec:meanfieldpropagation}) wave-packet propagation methods. Initial conditions are discussed in Sec.~\ref{InitialState}. In Sec.~\ref{sec:Results} we calculate the free-exciton properties of our model (\ref{sec:FreeIX}), while numerical details of wave-packet propagation are discussed in Sec.~\ref{sec:WavePacketPropagation}. Results are summarized for scattering potentials which are weak (Sec.~\ref{sec:WeakPotential}) or strong (in Sec.~\ref{sec:StrongPotential}) with respect to internal excitations, and for a single-slit geometry (Sec.~\ref{sec:SingleSlit}). Section \ref{sec:Conclusions} discusses, in particular, the predictivity of the different approaches. A formal derivation of the mean-field propagation scheme is provided in the Supplemental Material.\cite{SupplMat_IX}

\section{Theoretical approach}
\label{sec:theory}

\subsection{The electron-hole Hamiltonian}
\label{sec:model}

\begin{figure}
\includegraphics[trim={0 0 0 2.5cm}, clip, width=\columnwidth]{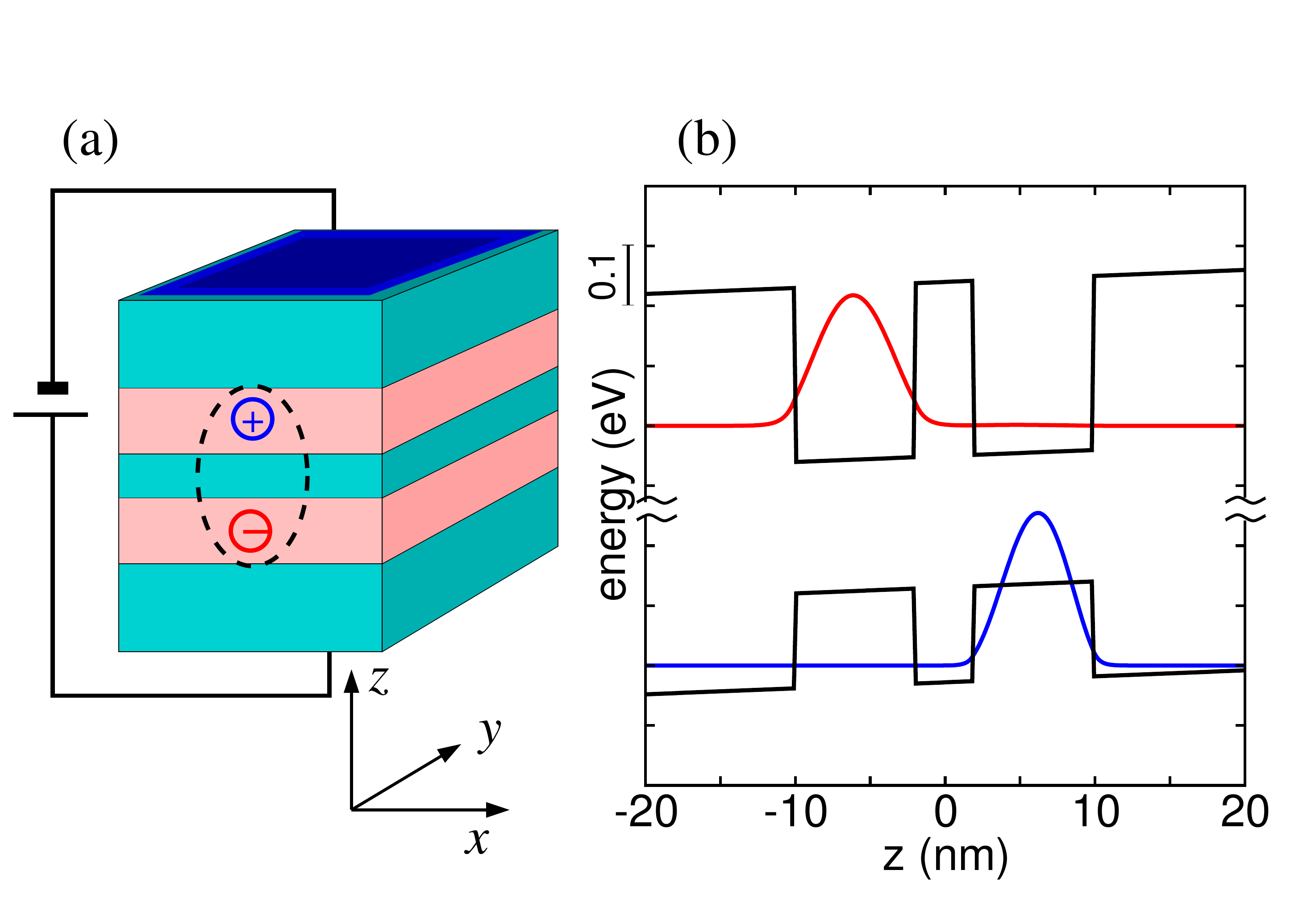}
\caption{(a) Sketch of the CQW heterostructure. Two symmetric GaAs quantum wells $8\,\mbox{nm}$ wide are embedded in Al$_x$Ga$_{1-x}$As matrix and separated by a $4\,\mbox{nm}$ wide barrier of the same material. The application of a vertical bias field separates the hole and the electron. (b) Calculated wave functions (square modulus) for the electron (red) and the hole (blue), and potential profile along the growth direction, for a Al$_x$Ga$_{1-x}$As barriers with $x=0.33$ and $F = 1\,\,\mbox{mV/nm}$. Material parameters are given in Tab.~\protect\ref{tab:parameters}.\label{zaxiswf}}
\end{figure}

Our reference system is sketched in Fig.~\ref{zaxiswf}(a). A symmetric GaAs CQW structure grown along $z$ is embedded in a Al$_x$Ga$_{1-x}$As matrix. A \textit{vertical} electric field $F_z$ along the growth direction separates electrons and holes in different layers.\cite{AndreakouPRB15}
While typical CQW confinement energies amount from tens to hundreds of meV, \textit{in-plane} $(xy)$ potential landscapes generated by metallic gates, as well as kinetic energies which can be impressed upon IXs, are in the meV range. Therefore, we factorize the in-plane and vertical component of the IX 3D wave function (here and throughout we use wave function in place of envelope function) as 
\begin{equation}
\Psi(\mathbf{r}_e,\mathbf{r}_h) \zeta_{e} (z_e) \zeta_{h} (z_h)\, ,\label{eq1_psitot}
\end{equation}
where $\mathbf{r}_i$ and $z_i$ are the 2D $xy$ coordinate and $z$ coordinate of the two particles, respectively. $ \zeta_{e} (z_e)$ and $ \zeta_{h} (z_h)$ are calculated explicitly for a given structure and field. $\zeta_{e}$
is calculated from the 1D effective-mass equation
\begin{equation}
H_{z_e} \zeta_e(z_e) = E_{z_e} \zeta_e(z_e)
\label{eveqz}
\end{equation}
with the Hamiltonian
\begin{equation}
 H_{z_e} =  -\frac{\hbar^2}{2} \frac{d}{d z_e}\left(\frac{1}{m_e^z(z_e)} \frac{d}{d z_e}\right) + W_e(z_e) - e z_e F \, , \label{Hamz}
\end{equation}
where $W_e(z_e)$ and $m_e^z(z_e)$ are the band edge and the material-dependent effective mass, respectively, of conduction electrons in the CQW structure. No spin-dependent term is considered.
Indeed, the validity of Eqs.~(\ref{eq1_psitot}) and (\ref{Hamz}) is limited to samples with narrow CQWs which are the typical heterostructures used in the experiments we are addressing to\cite{Gartner_APL06,Winbow_NL07,Winbow_JAP08,Leonard_APL12,Winbow_PRL11,High_Science11072008,Schinner_PRL13}. Here, in-plane Coulomb binding energy and scattering potentials are in the few meV range, while vertical confinement energies is at least one order of magnitude larger.

Equation (\ref{eveqz}) with the Hamiltonian (\ref{Hamz}) is called a Ben-Daniel Duke problem.\cite{Bastard_BOOK88} 
$E_{z_e}$ and $ \zeta_e(z_e)$ are obtained from (\ref{eveqz}) on a homogeneous real-space grid by a finite difference approach, taking into account the material-dependent effective mass. 
$E_{z_h}$ and $\zeta_{h}$ are computed similarly, with parameters appropriate to the valence band electrons. In this 
case $m_{h}^z$ is the effective mass given by the heavy-hole diagonal mass tensor, related to the Luttinger's parameters,\citep{VurgaftmanJAP01} $\gamma_1$ and $\gamma_2$, by
\begin{eqnarray}
m_{h}^z = (\gamma_1 - 2\gamma_2)^{-1}.
\end{eqnarray}
An example of these calculations is reported in Fig.~\ref{zaxiswf}(b) showing that the two carriers are well localized in either wells by the external bias. This justifies the separability of the electron and hole wave functions in the $z$ direction assumed in Eq.~(\ref{eq1_psitot}).

To simulate the electron-hole quantum dynamics we need to include accurately the mutual interaction. While in a ideal 2D electron system this is described by the bare $r^{-1}$ Coulomb potential, in a typical CQW heterostructure carriers are delocalized in the wells over a length which is comparable to the carrier separation, given by the effective Bohr radius. Therefore, the effective interaction is modified at short range, up to a distance comparable to the well width. To account for this effect, we consider electrons and holes in the ground state of the confinement potential (a reasonable assumption due to the large energy gaps in the growth direction) and we consider the \textit{effective 2D interaction} $U_C$ as the mean value of the Coulomb interaction over the wave functions in the growth direction,
\begin{equation}
\begin{split}
U_C(r) &= \langle \zeta_e \zeta_h |U_C^{3D}| \zeta_e \zeta_h \rangle \equiv \iint dz_e dz_h \times \\
& |\zeta_e(z_e)|^2 |\zeta_h(z_h)|^2 \left[ - \frac{e^2}{4\pi\epsilon_0\epsilon_r}\frac{1}{\sqrt{r^2 + (z_e-z_h)^2}} \right] .
\end{split}
\label{eq_U_C}
\end{equation}
Here, $r$ is the electron-hole in-plane distance and $\epsilon_r$ the relative permittivity of the well material.

Therefore, the full 3D problem has been mapped into an effective 2D model. 
The two-body wave function $\Psi$ is then propagated in time according to the 2D Hamiltonian 
\begin{equation}
H = H_0 + U_{\mbox{\scriptsize ext}}  \, ,
\end{equation}
where
\begin{equation}
H_0 = -\frac{\hbar^2}{2 m_e} \nabla_{\mathbf{r}_e}^2 -\frac{\hbar^2}{2 m_h^\parallel} \nabla_{\mathbf{r}_h}^2 + U_C(|\mathbf{r}_e-\mathbf{r}_h|)
\end{equation}
is the \emph{free IX Hamiltonian}. Here, $m_h^\parallel$ is the in-plane component of the strongly anisotropic  mass tensor of GaAs,
\begin{equation}
m_h^\parallel = (\gamma_1+\gamma_2)^{-1}\, .
\end{equation}

It is actually numerically convenient and more transparent to work in the CM and relative coordinate system
\begin{eqnarray}
\mathbf{R} & = & (m_e \mathbf{r}_e + m_h^\parallel\mathbf{r}_h)/M \,, \\
\mathbf{r} & = & \mathbf{r}_e - \mathbf{r}_h \,,
\end{eqnarray}
where $M= m_e + m_h^\parallel$ is the in-plane exciton mass and $m= [m_e^{-1}+(m_h^\parallel)^{-1}]^{-1}$ 
is the in-plane reduced effective mass. In this representation, the free IX Hamiltonian separates as
\begin{eqnarray}
H_{0}(\mathbf{R},\mathbf{r}) & = & H_{\mbox{\scriptsize CM}}(\mathbf{R}) + H_{\mbox{\scriptsize rel}}(\mathbf{r}) \,, \label{H0tot}\\
H_{\mbox{\scriptsize CM}}(\mathbf{R}) & \equiv & -\frac{\hbar^2}{2 M} \nabla_{\mathbf{R}}^2  \,, \\
H_{\mbox{\scriptsize rel}}(\mathbf{r}) & \equiv & -\frac{\hbar^2}{2 m} \nabla_{\mathbf{r}}^2  + U_C(r)  \,, \label{H0rel}
\end{eqnarray}
and the free in-plane wave function can be factorized as 
\begin{eqnarray}
\Psi(\mathbf{R},\mathbf{r}) = \chi(\mathbf{R})\phi(\mathbf{r}) \,.
\label{eq:factoriation}
\end{eqnarray}
As we shall discuss later, the CM energy is typically of the order of tenths of meV, much less than the internal excitation energy, which is of the order of several meV. Therefore, the grid parameters related to CM and relative dynamics are quite different and can be optimized separately. On the contrary, using electron and hole coordinates requires to use (almost) the same grid, and the same accuracy would be reached at a greater computational cost.\citep{GrasselliJCP15}

In this representation, the inclusion of one-body external potentials, $U_{\mbox{\scriptsize ext}} = U_e(\mathbf{r}_e) + U_h(\mathbf{r}_h) = U_{\mbox{\scriptsize ext}}(\mathbf{R}, \mathbf{r})$ in the full Hamiltonian couples the CM and relative coordinates, and removes the separability of the wave function, Eq.~(\ref{eq:factoriation}). Therefore, to propagate $\Psi(\mathbf{R}, \mathbf{r};t)$ we need to deal with a four DoFs propagation scheme.
We neglect the interaction of IXs with environmental degrees of freedom, like phonons of the semiconductor lattice.  In typical experiments $E_\mathrm{CM}$ is in the tenths of meV range, well below the optical phonon energy in GaAs, while low temperature strongly suppresses acoustic phonon population. In our simulation, the scattering time is in the order of tens of picoseconds, well below the LA-phonon assisted relaxation of IXs in these devices, which is in the nanoseconds range.\citep{Butov_JPCM04}

\subsection{Full numerical propagation}
\label{sec:fullpropagation}

The quantum propagation of the IX is obtained through the numerical application of the evolution operator, $\mathcal{U}(t+\Delta_t;t)$, between two consecutive times $t$ and $t+\Delta_t$ as
\begin{eqnarray}
\Psi(\mathbf{R},\mathbf{r};t+\Delta_t) = \mathcal{U}(t+\Delta_t;t) \Psi(\mathbf{R},\mathbf{r};t) .
\end{eqnarray}
Our numerical solution relies on the Fourier split step (FSS) approach.\cite{Leforestier_JCP91,Castro_JCP04} This unitary method, numerically exact as $\Delta_t\rightarrow0$, is based on the Suzuki-Trotter factorization\cite{Suzuki_PJAB93} of the evolution operator in the product of two exponential operators, containing the kinetic or the potential operators, respectively, each diagonal either in direct or in reciprocal space (see Appendix in Ref.~\onlinecite{GrasselliJCP15} for details):
\begin{eqnarray}
\begin{split}
&\mathcal{U}(t + \Delta_t; t) =\\
&= e^{ -\frac{i}{\hbar} U_{\mathrm{tot}} \frac{\Delta_t}{2} }
          e^{ -\frac{i}{\hbar} T  \Delta_t } e^{ -\frac{i}{\hbar} U_{\mathrm{tot}}  \frac{\Delta_t}{2}} + \mathcal{O}(\Delta_t^3).\label{eq:SuzTrot}
\end{split}
\end{eqnarray}
where $T = \mathbf{P}^2/(2M) + \mathbf{p}^2/(2m)$ and $U_{\mathrm{tot}} = U_C + U_\mathrm{ext}$ are the total kinetic and potential energy operators of the system ($\mathbf{P}$ and $\mathbf{p}$ are CM and relative in plane linear momenta, respectively).  
Hence, at each time step the IX wave function must be switched from position to momentum representation, and \textit{vice versa}, through Fourier transformation, $\mathcal{F}$, according to
\begin{eqnarray}
\begin{split}
\Psi(\mathbf{R},\mathbf{r}; & t + \Delta_t) =e^{ -\frac{i}{\hbar} U_{\mathrm{tot}} \frac{\Delta_t}{2} }\times \\
& \mathcal{F}^{-1}\left\lbrace 
          e^{ -\frac{i}{\hbar} T  \Delta_t } \mathcal{F} \left[ e^{ -\frac{i}{\hbar} U_{\mathrm{tot}}  \frac{\Delta_t}{2}}\Psi(\mathbf{R},\mathbf{r};t) \right]\right\rbrace.\label{eq:FourierSS}
\end{split}
\end{eqnarray}
The use of the Fast Fourier Transform (FFT) algorithm, therefore, results in high computational efficiency with respect to other methods, particularly those based on finite difference discretized Hamiltonian, as the Crank-Nicolson method. 

The coupling between CM and relative DoFs introduced by the one-body external potentials $U_{\mbox{\scriptsize ext}}(\mathbf{R},\mathbf{r})$ requires the numerically full propagation of all four DoFs of the IX for a sufficiently long time for the scattering process to conclude on a sufficiently dense and extended real-space grid. For the energy scales into play here, this turns out to be a demanding numerical task, both to store the complex-valued IX wave function in memory, and to numerically compute the application of the evolution operators to it. A code exploiting massive parallelization has been developed to cope with these issues. The four-dimensional domain is discretized in a grid of about $4295\times10^6$ points. Typically, the
computation of a single $40$~fs time step takes $10.5$~seconds, and one Gb of memory per core is used on a 256-core run (with 16 cores in two 2.4 GHz Haswell Xeon processors per shared-memory node), corresponding to a speedup of about 43 with
respect to a serial run.

\subsection{Mean-field propagation}
\label{sec:meanfieldpropagation}

In semiconductor physics, when dealing with excitons in weak potentials, it is often justified to apply the so-called \textit{rigid exciton model} (RIX),\cite{Zimmermann_PAC97} which consists in the assumption that the IX is frozen into its relative motion ground state, $\phi_0(\mathbf{r})$; hence, only the quantum evolution of the CM component, $\chi(\mathbf{R})$, is taken into account. The internal DoFs are integrated out, leading to an effective potential
\begin{eqnarray}
U_{\mbox{\scriptsize eff}}(\mathbf{R}) = \iint d\mathbf{r} |\phi_0(\mathbf{r})|^2 U_{\mbox{\scriptsize ext}}(\mathbf{R},\mathbf{r}). \label{eqRIXeffpot}
\end{eqnarray}
Therefore, the IX moves as a rigid object with coordinates $\mathbf{R}$, its dynamics being determined by the external potential averaged on the relative-motion ground state. For example, IX wave function localization in weak traps can be calculated in this way.\cite{Hohenester_APL04} 

The CM effective evolution operator
\begin{eqnarray}
\mathcal{U}_{\mbox{\scriptsize eff}} (t+\Delta_t;t) \equiv \exp\left\lbrace-\frac{i}{\hbar}\left[ \frac{\mathbf{P}^2}{2M} + U_{\mbox{\scriptsize eff}}(\mathbf{R}) \right]\Delta_t \right\rbrace
\end{eqnarray}
is then applied to the CM wave function, $\chi(\mathbf{R},t)$, and the numerical evolution can be obtained again by the FSS method. 
Obviously, the RIX model requires a much lower computational effort with respect to full propagation, since only the two DoFs wave function $\chi(\mathbf{R})$ needs to be propagated, which can be readily obtained on a standard personal computer. Clearly, every effect of the internal dynamics is neglected by the RIX approximation.  

The RIX model is the lowest order example of a more general mean-field strategy, also known as the \textit{time dependent Hartree} (TDH) method in atomic and molecular scattering.\cite{McLachlan64,BeckTDH2000} Within this approach, at each time $t$ the global wave function of the composite object is assumed to be factorized into a CM and a relative wave function, 
\begin{eqnarray}
\Psi(\mathbf{R},\mathbf{r};t) = \chi(\mathbf{R};t)\phi(\mathbf{r};t) .
\end{eqnarray}
The evolution of $\chi(\mathbf{R};t)$ and $\phi(\mathbf{r};t)$ is determined by an effective potential representing the expectation value of the external potential on the relative and CM wave function, respectively, \textit{at that specific time} $t$, i.e. (see Ref.~\onlinecite{McLachlan64}, and Suppl. Mat.\cite{SupplMat_IX})
\begin{equation}
\begin{split}
\chi(\mathbf{R};t+\Delta_t) &= \exp \left\lbrace - \frac{i}{\hbar} \int_t^{t+\Delta_t}dt' \times \right.\\
& \left. \left[ \frac{\mathbf{P}^2}{2M} +  U_{\mbox{\scriptsize eff}}(\mathbf{R};t')   \right] \right\rbrace \chi(\mathbf{R};t) 
\end{split} \label{eqTDHCM}
\end{equation}
and
\begin{equation}
\begin{split}
\phi(\mathbf{r};t+\Delta_t) &= \exp \left\lbrace -\frac{i}{\hbar} \int_t^{t+\Delta_t} dt' \times \right.\\
& \left. \left[ \frac{\mathbf{p}^2}{2m} + U_{C}(\mathbf{r}) + u_{\mbox{\scriptsize eff}}(\mathbf{r};t') \right] \right\rbrace \phi(\mathbf{r};t) 
\end{split} \label{eqTDHrel}
\end{equation}
with
\begin{eqnarray}
U_{\mbox{\scriptsize eff}}(\mathbf{R};t) \equiv \int d\mathbf{r} |\phi(\mathbf{r};t)|^2 U_{\mbox{\scriptsize ext}}(\mathbf{R},\mathbf{r})  
\label{effTHDpotss}
\\
u_{\mbox{\scriptsize eff}}(\mathbf{r};t) \equiv \int d\mathbf{R} |\chi(\mathbf{R};t)|^2 U_{\mbox{\scriptsize ext}}(\mathbf{R},\mathbf{r}) \label{effTHDpots}
\end{eqnarray}

This is a mean-field model, since the evolution of the CM wave function is determined by the average field generated by the relative wave function, and \textit{vice versa}, at each time step. Again, the numerical propagation is performed using the FSS method. The FSS algorithm needs to be applied twice at each time step, independently of the CM and relative wave functions, the two evolutions being coupled through the effective potentials, Eqs.~(\ref{effTHDpotss}), and (\ref{effTHDpots}). Clearly, the RIX model consists in assuming a \emph{rigid} $\phi(\mathbf{r};t) = \phi_0(\mathbf{r})$, and only the $\chi(\mathbf{R};t)$ component needs to be propagated.

Note that, even for a stationary external potential $U_{\mbox{\scriptsize ext}}(\mathbf{R},\mathbf{r})$, the mean-field propagation method requires to evolve the two components of the wave function under  \emph{time-dependent} potentials. This increases substantially the computational cost of the simulation with respect to the RIX model, since the propagator needs to be calculated at each time step rather than only once. However, the TDH approach is still far less demanding than the full evolution. 

\subsection{Initial state}
\label{InitialState}

In order to start a time dependent simulation, we need to choose a proper initial state. 
In typical CQW systems, an IX thermalizes and relaxes to the ground state $\phi_0(\mathbf{r})$ of the free exciton Hamiltonian $H_0$ within nanoseconds from photogeneration, due to active scattering mechanisms (acoustic phonons). 
We do not consider in our simulations this transient, which is short compared to the IX photo-recombination lifetime.\citep{Butov_JPCM04}
Therefore we initialize the IX wave function as
\begin{eqnarray}
\Psi(\mathbf{R},\mathbf{r};t=0) = \chi(\mathbf{R})\phi_0(\mathbf{r}) \,.
\end{eqnarray}
The CM wave function is chosen as the minimum uncertainty wave-packet
\begin{eqnarray}
\chi(\mathbf{R}) =& \left(\frac{1}{2\pi\sigma_X\sigma_Y}\right)^{1/2} \exp(i\mathbf{K}_0\cdot \mathbf{R}) \times \nonumber \\
& \exp\left[-\frac{(X-X_0)^2}{4\sigma_X^2}\right]\exp\left[-\frac{(Y-Y_0)^2}{4\sigma_Y^2}\right],
\end{eqnarray}
centered at the initial CM position $(X_0,Y_0)$, having widths $\sigma_X$ and $\sigma_Y$, and propagating with an average CM wave vector
\begin{eqnarray}
\mathbf{K}_0 \equiv \sqrt{\frac{2M E_{\mbox{\scriptsize CM}}}{\hbar^2}}(\sin\theta, \cos\theta) ,
\end{eqnarray}
where $E_{\mbox{\scriptsize CM}} $ is the most probable CM kinetic energy at $t=0$ and $\theta$ identifies the initial propagation direction with respect to a properly defined normal incidence direction. In the following simulation we take, as initial dispersion, $\sigma_X=\sigma_Y=80$~nm unless otherwise specified. This value is of the same order of magnitude of the confinement length of IX traps~\cite{Schinner_PRL13}. This is a reasonable compromise between a sufficiently narrow momentum distribution (hence a delocalized exciton) and a spatially localized IX.
The CM momentum can be controlled, e.g., through the application of acceleration ramps. Localization of the initial state can be controlled by electrostatic traps. See Refs.~\onlinecite{Gartner_APL06,Winbow_NL07,Winbow_JAP08,Leonard_APL12,Winbow_PRL11,High_Science11072008,Schinner_PRL13,ViolanteNJP14} for typical parameters, comparable to those used in our simulations. Note that, while the CM energy clearly affects transmission and reflection coefficients, due to the linearity of the equations the momentum distribution does not affect much the dynamics, as long as the momentum dispersion is not too broad.

We remark that, even if the initial state is factorized, in the full propagation the wave function is correlated, evolving under the influence of both the electron-hole interaction and the external potential, with no specific \textit{a priori} decomposition. On the contrary, factorization for the IX wave function is assumed in the RIX and TDH approximations at any intermediate time $t$. 

\section{Results}
\label{sec:Results}

In the following we investigate the IX dynamics in a 8nm/4nm/8nm GaAs/Al$_{0.33}$Ga$_{0.67}$As/GaAs CQW system.\citep{Butov_NatureA02,Butov_NatureB02,Butov_JPCM04}
Band parameters are indicated in Tab.~\ref{tab:parameters}. An homogeneous (i.e. constant along the $xy$ planes) electric field $F =1 \mbox{mV/nm}$, generated by the application of a bias voltage between the top and the back gates of the sample, is assumed (see Fig.~1).

\begin{table}
\begin{tabular}{lcc} 
           & GaAs       & Al$_{0.33}$Ga$_{0.67}$As \\
 \hline
 $m_e/m_0$ (Ref.~\onlinecite{Levinshtein96}) & 0.067 & 0.094\\
 $\gamma_1,\gamma_2$ (Ref.~\onlinecite{VurgaftmanJAP01}) & {$6.98, \, 2.06$} & \\
 Relative permittivity (Ref.~\onlinecite{Levinshtein96}) & {12.9} & \\
 Valence band offset (Ref.~\onlinecite{YuSSP92}) & \multicolumn{2}{c}{0.158 eV} \\
 Conduction band offset (Ref.~\onlinecite{BosioPRB88}) & \multicolumn{2}{c}{0.291 eV} \\
  \hline
\end{tabular}
\caption{Band parameters adopted in the simulations.}
\label{tab:parameters}
\end{table}

\subsection{Free IX: effective Coulomb interaction and relative motion eigenvalue problem}
\label{sec:FreeIX}

In Fig.~\ref{zaxiswf}(b) we show the square modulus of the calculated electron and hole wave function components ($\zeta_{e}$, $\zeta_{e}$), together with the band profile along the growth axis of the heterostructure. Clearly, the electric field localizes the electron and the hole in different layers. 

\begin{figure}
\includegraphics[trim={0 0 0 4cm}, clip, width=\columnwidth]{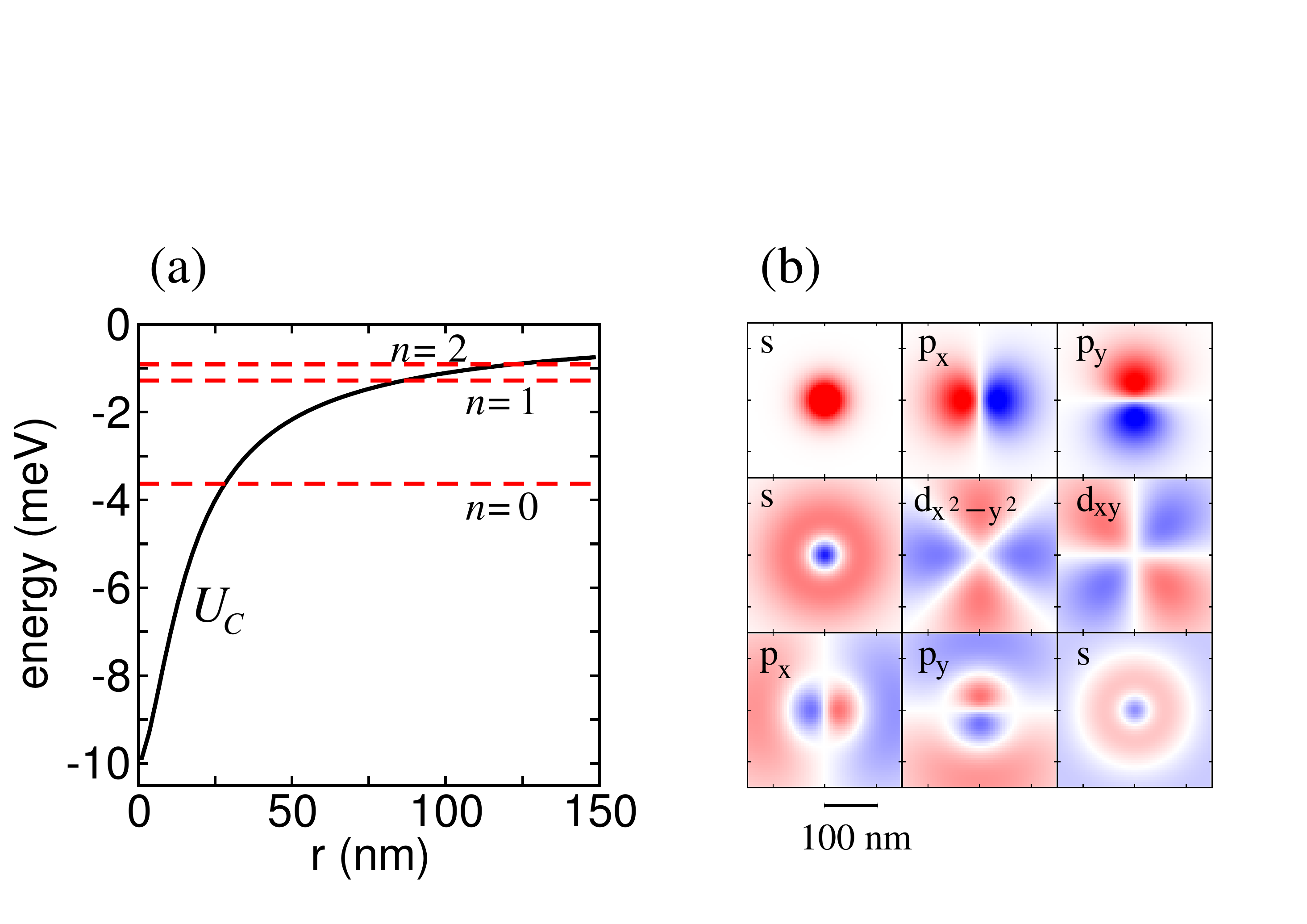}
\caption{(a) Solid line: Calculated effective electron-hole potential $U_C(r)$ used in the simulations. Dashed lines: lowest bound IX energy levels. (b) wave functions of the lowest bound electron-hole eigenstates. The scale is arbitrary. Blue and red regions denote opposite sign values. See also Tab.~\ref{tab:energies}}
\label{rel_Coul_states}
\end{figure}

$\zeta_{e}, \zeta_{h}$ are  used to compute the effective electron-hole interaction $U_C(r)$ (see Eq.~(\ref{eq_U_C})) along the CQW planes, shown in Fig.~\ref{rel_Coul_states}(a). At large electron-hole distances $U_C$ amounts to the bare Coulomb interaction. At small distances, the Coulomb divergence is removed due to the separation of the electron and the hole in different layers. Bound electron-hole states and energy levels $\mathcal{E}_n$ are calculated from $H_{\mbox{\scriptsize rel}}$ with a finite difference approach on a 2D uniform square grid. The grid density is the same as for the wave packet evolution (see Sec.~\ref{sec:WavePacketPropagation}). The lowest bound states are shown in Fig.~\ref{rel_Coul_states}(a). 
Energies and degeneracies of the lowest states are reported in Tab.~\ref{tab:energies}, together with their symmetries, as deduced from Figure~\ref{rel_Coul_states}(b), which shows the wave functions $\phi_n(\mathbf{r})$ of the nine lowest states. 

\begin{table}
\begin{tabular}{cccc}
 $n$ & symmetry & degeneracy & $\mathcal{E}_n$ (meV) \\
  \hline
  0   & $s$ & 1 & -3.63 \\
  1   & $p$ & 2 & -1.28 \\
  2   & $s$ & 1 &  -0.91 \\
  3   & $d$ & 2 &  -0.53 \\
  4   & $p$ & 2 &  -0.49 \\
  5   & $s$ & 1 &  -0.42 \\
  \hline
\end{tabular}
\caption{Lowest energy levels of the free IX relative motion Hamiltonian, Eq.~(\ref{H0rel}).}
\label{tab:energies}
\end{table}

\subsection{Wave packet propagation}
\label{sec:WavePacketPropagation}

Unless differently indicated, the simulations described below have been performed with the following parameters: a total simulation time of $60\,\mbox{ps}$ and  a time step $\Delta_t =$ 40 fs; a spatial CM domain with $\mathbf{R}\equiv(X,Y)\in [-1.5\,\mu\mathrm{m}, 1.5\,\mu\mathrm{m}]^2$ and a grid point density of 0.17 points/nm along both directions; a spatial relative motion domain with $\mathbf{r}\equiv(x,y)\in [-0.15\,\mu\mathrm{m}, 0.15\,\mu\mathrm{m}]^2$ and a grid point density of 0.43 points/nm in both directions. 

These parameters satisfy the Nyquist criterion\cite{NR_2007} for the spatial frequency sampling far beyond the considered energy ranges. Another criterion in order to apply the FSS method with no ambiguity requires that the phase exponent $ U_{\mbox{\scriptsize ext}} \Delta_t/(2\pi\hbar) \ll 1$. If this condition is not fulfilled, the external potential strength is not well defined, because the class of $\mod(2\pi \hbar/\Delta_t)-$defined potentials has more than one element. In this case the method gives the same results with different potentials, which is physically inconsistent.

In order to estimate the IX localization during the time dependent simulations and, in the asymptotic times, the transmission and reflection probabilities, we define three regions of the CM space: (i) the \emph{reflection} region \textbf{A}, i.e., the subspace with vanishing external potential, $U_e=U_h=0$, where the IX wave function is initially localized;
(ii) the \emph{potential} region \textbf{B}; (iii) the \emph{transmission region} \textbf{C}, i.e., the subspace with vanishing external potential which can be reached from \textbf{A} only by crossing the potential region. 
Note that these regions refer to the CM DoFs only, while in the four DoF model the external potential $U_{\mbox{\scriptsize ext}}$ depends also on the relative coordinate $\mathbf r$. Thus, in order to compare consistently the different methods, we define the above regions according to the potential of Eq.~(\ref{eqRIXeffpot}). Specifically, we choose as boundaries of the region \textbf{B} the $Y$ positions where the external effective potential drops to 5\% of its maximum value. The related coefficients are then defined as integrals, over the corresponding regions 
\begin{itemize}
\item of the CM-part of the wave function square modulus, $|\chi(\mathbf{R};t)|^2$, for the mean field approximations;
\item of the CM marginal probability, 
\begin{equation}\label{eq:MarginalProbability}
\rho_{\mbox{\scriptsize CM}}(\mathbf{R};t) \equiv \int d\mathbf{r} |\Psi(\mathbf{R},\mathbf{r};t)|^2,
\end{equation} for the full propagation. 
\end{itemize}
These integrals are then normalized to the whole domain, so that the \textbf{A},\textbf{B},\textbf{C}-coefficients are defined in the interval $[0,1]$. An example of the three regions is showed in Fig.~\ref{weakpot}. Clearly, the three coefficients evolve in time. Coefficients \textbf{A} and \textbf{C} at asymptotic times set to a constant which are the reflection and transmission probabilities, respectively.

Below we shall investigate the time-dependent dynamics of a wave-packet, prepared as described in Sec.~\ref{InitialState}, scattering against two classes of potentials, a uniform, infinitely long barrier or well, and a barrier with a slit, which mimics the potential generated by a split gate, as typically realized in 2D heterostructures. A few comments are in order:
\begin{itemize}
\item[-] The bias and/or gating potential drops are much smaller than the confinement energies in the quantum wells of the structure. Therefore, wave functions $\zeta_{e}$, $\zeta_{h}$  are not distorted by \emph{local} variations of the external potential, which is thus uninfluential on the vertical localization of the electron and the hole. In other words, the effective interaction $U_C$ can be considered independent of space ($\mathbf{R}$) and time;
\item[-] In CQW systems $U_e$ and $U_h$ are in general different, with opposite sign. A metallic gate on top of the structure, for example, generates an electrostatic potential which is opposite for electrons and holes, and it is slightly different in strength between the two layers, due to the different distance from the gate. A simple capacitor model\cite{GrasselliJCP15} shows that the difference is typically of a few meV and comparable to the generated in-plane voltage drop.\cite{note_gating} 
\item[-] We shall investigate external potentials which are short ranged and vanish exactly outside the scattering region. According to the energy conservation law, excitations to higher internal IX levels, up to the dissociation threshold, are allowed only \textit{inside} the scattering region \textbf{B}; in the asymptotic regions (\textbf{A}, \textbf{C}), the CM energy (which in our simulations is always much smaller than the lowest internal excitation threshold, see Tab.~\ref{tab:energies}) is not sufficient to excite the internal dynamics, and the IX can only be transmitted/reflected in the internal ground state. 
\end{itemize} 

Two regimes can be identified, with the strength of the scattering potential being weak or strong with respect to the internal motion excitation energies. Below we shall investigate separately these two regimes for uniform, infinitely long well/barriers potentials, chosen separately for the two particles, but with a common width $L_y$,
\begin{equation}
U_{\mbox{\scriptsize ext}} = 
\left\{\begin{array}{ll}
U_{e,0} + U_{h,0} & \mbox{if} \; 0\leq y \leq L_y \\
0 & \mbox{otherwise}
\end{array}\right.
\label{eq:potential}
\end{equation} 

\subsection{Weak external potential}
\label{sec:WeakPotential}

\begin{figure}
\includegraphics[trim={0 0 0 8cm}, clip, width=\columnwidth]{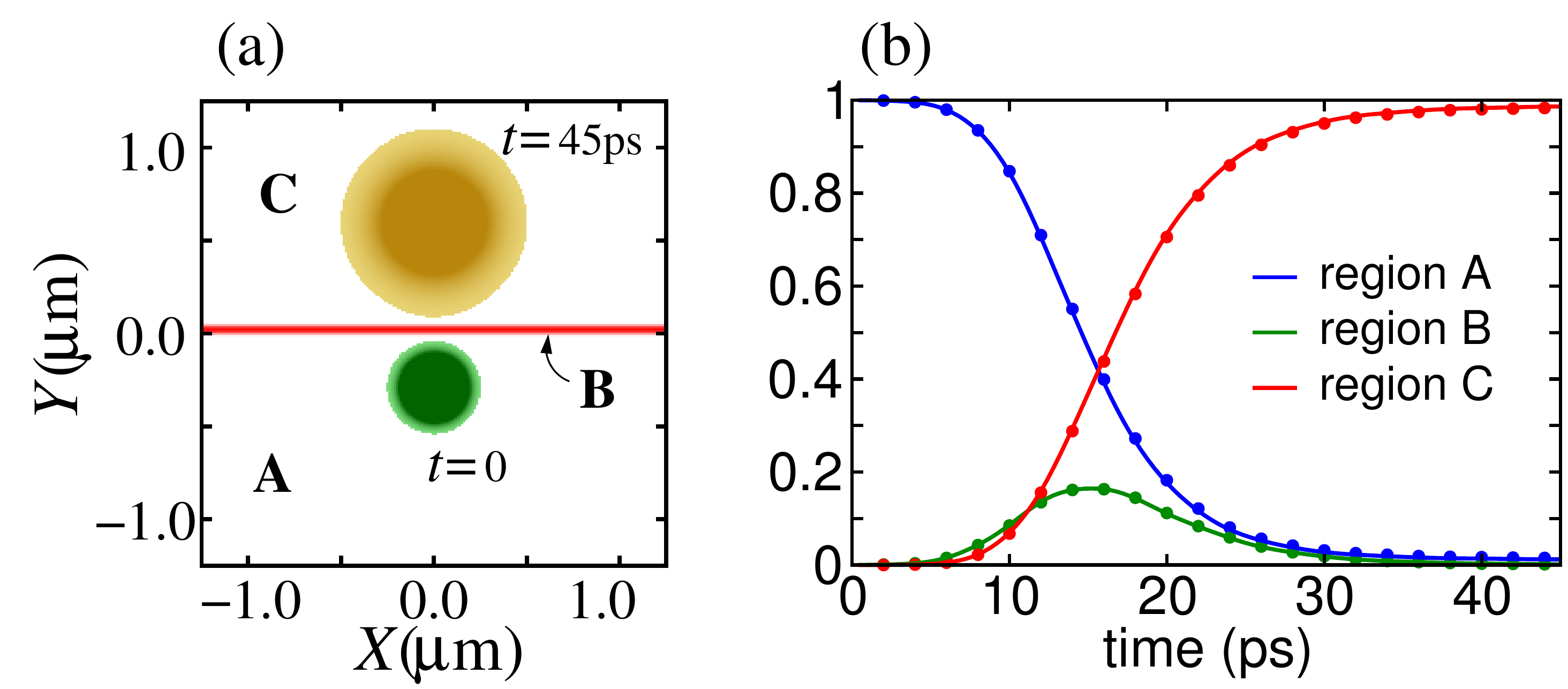}
\caption{(a) CM domain with hole potential well (see Eq.~(\ref{eq:potential}) $U_{h,0}=-1 \,\mbox{meV}$, $U_{e,0}=0$, $L_y = 40 \,\mbox{nm}$, and $E_{\mbox{\scriptsize CM}}=0.2 \,\mbox{meV}$. The reflection (\textbf{A}), external potential (\textbf{B}), and transmission (\textbf{C}) regions are indicated. The CM marginal probability (see main text) is shown at two different times, $t=0$ and $t=45\,\mbox{ps}$. The initial position is set at $(X_0,Y_0)=(0,-300)\,\mbox{nm}$. A short animation of this case is presented in the Supplemental Material\cite{SupplMat_IX} (b) Evolution of the \textbf{A},\textbf{B},\textbf{C}-coefficients with time. Solid line: full propagation. Dots: RIX approximations.}\label{weakpot}
\end{figure}

In Fig.~\ref{weakpot} we show a simulation for a weak potential well for the hole, $U_{h,0} = -1.0\,\mbox{meV}, U_{e,0}=0$,  and $L_y=40\,\mbox{nm}$. Note that the external potential strength is smaller than the lowest internal excitation energy, $\mathcal{E}_1-\mathcal{E}_0 = 2.35\,\mbox{meV}$ (see Tab.~\ref{tab:energies}). The IX is initialized with a kinetic energy $E_{\mbox{\scriptsize CM}} = 0.2 \,\mbox{meV}$ with normal incidence ($\theta=0$) to the well. 

Figure \ref{weakpot}(a) shows that the IX is completely transmitted as a bound state in the internal ground state, as in a single-particle scattering. This is confirmed in Fig.~\ref{weakpot}(b) where we plot the time evolution of the \textbf{A},\textbf{B},\textbf{C}-coefficients, showing that scattering is over in 
$\sim 20\,\mbox{ps}$, and the wave-packet is completely transmitted. Furthermore, the full propagation and the RIX and TDH propagations give indistinguishable results (we did not plot the TDH calculation for clarity). This proves that in this regime \textit{i)} the wave function can be factorized into the product $ \Psi(\mathbf{R},\mathbf{r};t) = \chi(\mathbf{R};t)\phi(\mathbf{r};t) $ and \textit{ii)} it remains in the ground state $\phi(\mathbf{r};t) = \phi_0(r)e^{-i\mathcal{E}_0 t/\hbar}$ of the internal DoFs, the fundamental assumption in the RIX approximation, during the \textit{whole} propagation.

\subsection{Strong external potential}
\label{sec:StrongPotential}

We next investigate scattering of an IX against external potentials with an energy scale comparable to the IX internal excitations.

\subsubsection{Electron well}\label{Sec:elec_well}

\begin{figure}
\includegraphics[width=\columnwidth]{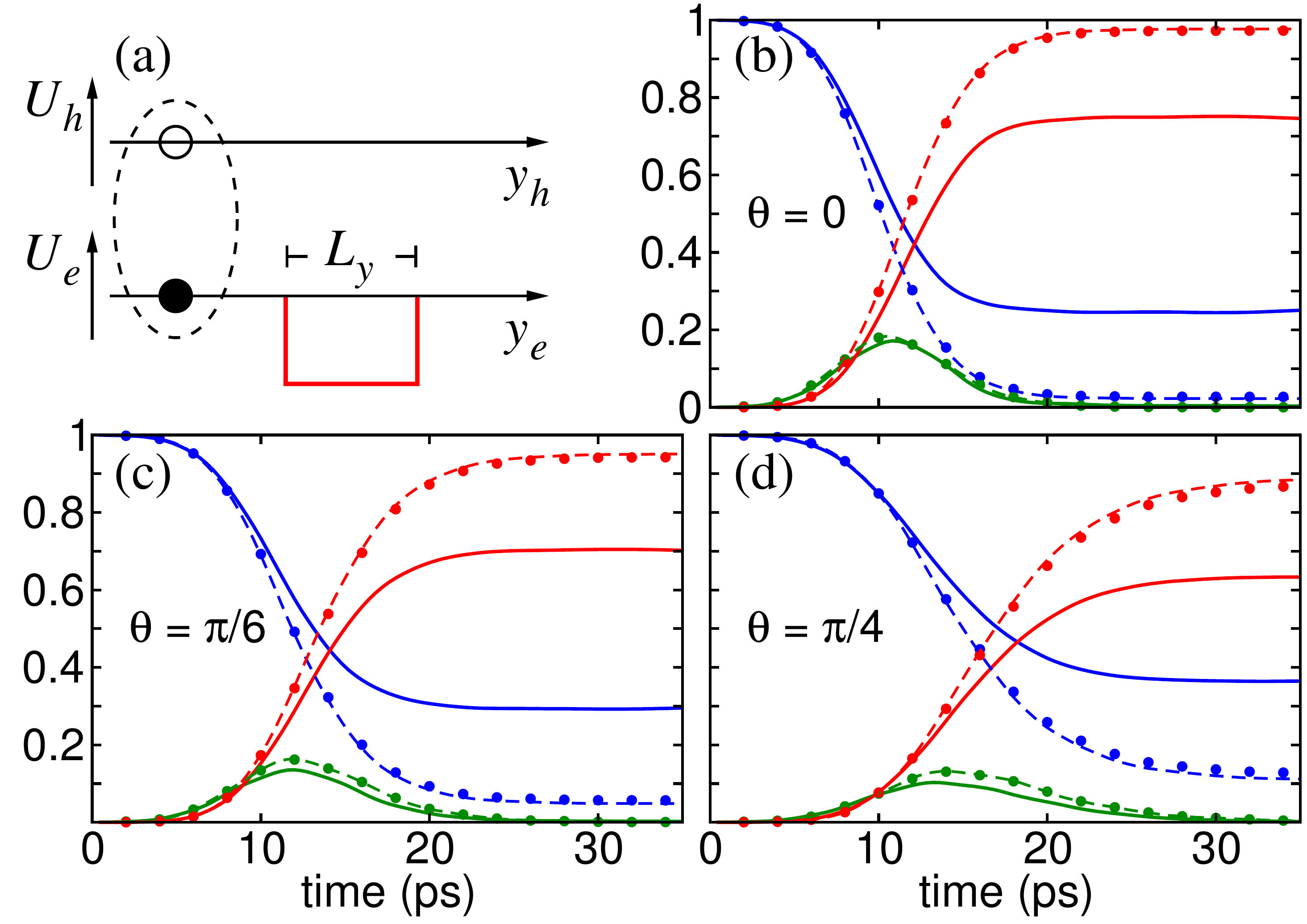}
\caption{Scattering in an electron well with $U_{e,0}=-3 \,\mbox{meV}$, $U_{h,0}=0$, $L_y = 40 \,\mbox{nm}$. (a) Sketch of the external potential. (b-d) Time evolution of the \textbf{A},\textbf{B},\textbf{C}-coefficients (line colors as in Fig.~3) at normal incidence $\theta=0$, $\theta=\pi/6$, $\theta=\pi/4$, as indicated. Dots: RIX approximation. Dashed lines: TDH approximation. Solid line: full propagation.}
\label{elec_well}
\end{figure}

We first consider scattering of a IX with an external potential consisting of a square well applied to the electron,   $U_{e,0} = -3.0\, \mathrm{meV}$, $U_{h,0}=0$, and $L_y=40\,\mbox{nm}$ (see Fig.~\ref{elec_well}(a)). The initial CM kinetic energy of the IX is set to $E_{\mbox{\scriptsize CM}}=0.4 \,\mbox{meV}$.
Now the external potential intensity is slightly below the dissociation energy, $-\mathcal{E}_0 = 3.63 \,\mbox{meV}$, but it is sufficient to excite the IX to higher energy internal states. 
The dissociation phenomena, not possible here due to energy conservation, has been analyzed elsewhere for a simpler 1D geometry\cite{GrasselliJCP15}.

\begin{figure}
\includegraphics[trim={0 0 2.5cm 0cm}, clip, width=\columnwidth]{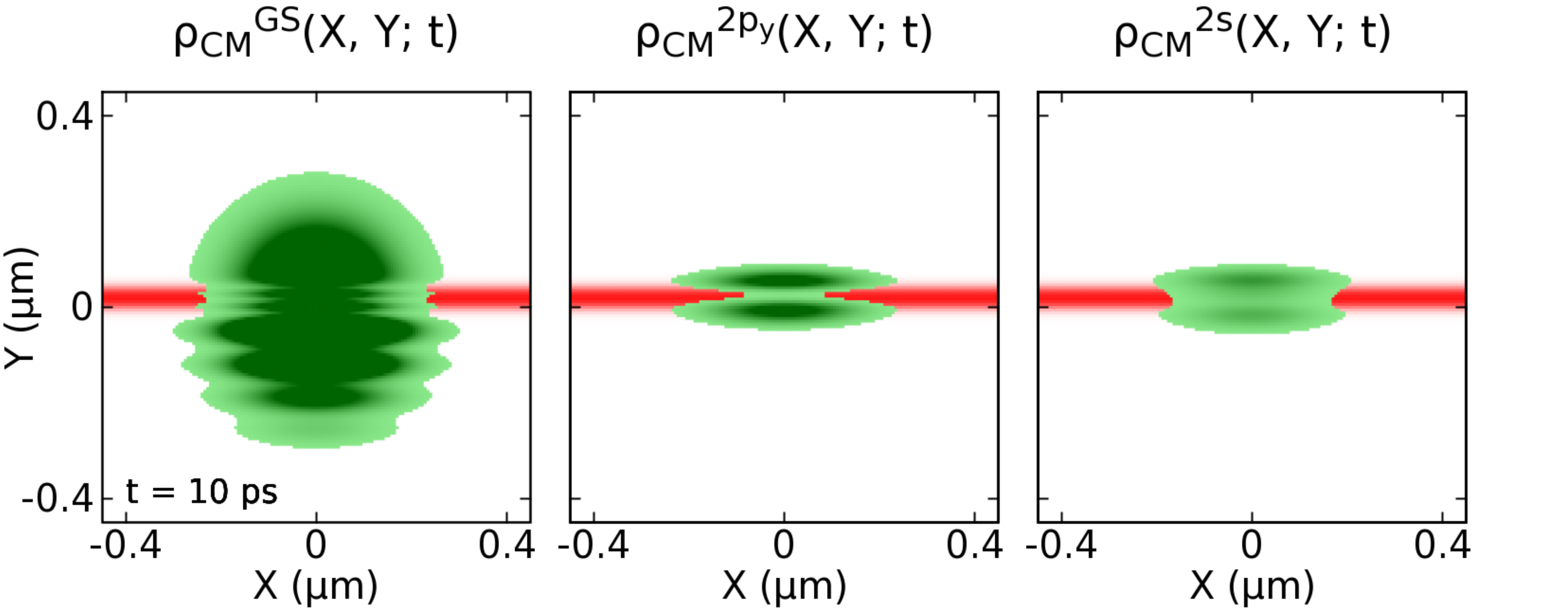}
\caption{Projections $\rho_\mathrm{CM}^j(\mathbf{R}; t)$, of the total wave function, on the ground state (left), on the first $p_y$ excited state (center) and on the $2s$ state (right) of the relative motion Hamiltonian, for the normal incidence scattering onto an electron well with $U_{e,0}=-3 \,\mbox{meV}$, $U_{h,0}=0$, $L_y = 40 \,\mbox{nm}$. A short animation is presented in the Supplemental Material.\cite{SupplMat_IX} }
\label{elec_well_proj}
\end{figure}

To highlight the role of internal excitations, in Fig.~\ref{elec_well_proj} we show the projections 
\begin{equation}
\rho_\mathrm{CM}^j(\mathbf{R}; t) = \left|\int d\mathbf{r} \phi_j(\mathbf{r}) \Psi(\mathbf{R},\mathbf{r};t)\right|^2
\end{equation}
on the first internal eigenstates $j=1s,2p_y,2s$ at $t=10$ ps. The $p_x$-state projection is not shown, since it cannot be excited due to potential symmetry reasons.
We see that the excitation of the internal DoF takes place especially at the edge of the well, where the change in potential energy is abrupt. $\rho_\mathrm{CM}^j(\mathbf{R}; t)$, for $j\neq 1s$, vanish as $t\gtrsim 20$ ps, i.e. when the scattering process is almost concluded.
  
In Fig.~\ref{elec_well}(b-d) we show the \textbf{A},\textbf{B},\textbf{C}-coefficients at normal incidence $\theta=0$, and $\theta=\pi/6$ and $\theta=\pi/4$ incidence. In all these cases, scattering is over in $\sim 15\div20\,\mbox{ps}$, the scattering time being larger for larger incident angles. The transmitted wave-packet and the amplitude in the potential region are smaller as the angle increases. This is in agreement with the lower momentum of the exciton in the direction normal to the potential well, \emph{if scattering is not at a resonance energy}.

In this regime, RIX or TDH substantially overestimate the transmission obtained from full propagation, while RIX and TDH give very similar results between each other. Note however, that the trend with incident angle is similar for the three methods and that in the potential region \textbf{C} all methods give very similar results. Therefore, even though the CM localization is similar during scattering, the internal DoFs have a strong effect on the transmission and reflection at asymptotic times.

We also verified by explicit simulations that transmission with $\theta >0$ coincides with that obtained at normal incidence but with a kinetic energy which corresponds to the normal component of CM wave vector, which for the present case is $0.3\,\mbox{meV}$ at $\theta=\pi/6$, and $0.2 \,\mbox{meV}$ at $\theta=\pi/4$. This holds true both in the full propagation and in the mean-field methods. This is clearly to be expected in the RIX approximation, which is effectively a one-particle problem, since the evolving CM wave function is separable into an $X$ and a $Y$ part in a translationally invariant potential. For full and TDH propagations, where the Coulomb interaction couples $x$ and $y$ directions, note that for an external potential which is invariant in one planar direction, say, with respect to $x_e, x_h$, the \emph{total} potential is invariant with respect to $X$, $U_C(x,y)+U_{\mbox{\scriptsize ext}}(Y,y)$. Therefore, the $X$ Fourier components of the full wave function is not scattered by the potential, and separates as 
$\Psi(\mathbf{R},\mathbf{r}) = e^{i K_X X} \Psi_{\perp}(Y,x,y)$. Accordingly, the dynamics corresponds to the free-particle one in the $X$ direction.

\subsubsection{Hole well}

\begin{figure}
\includegraphics[width=\columnwidth]{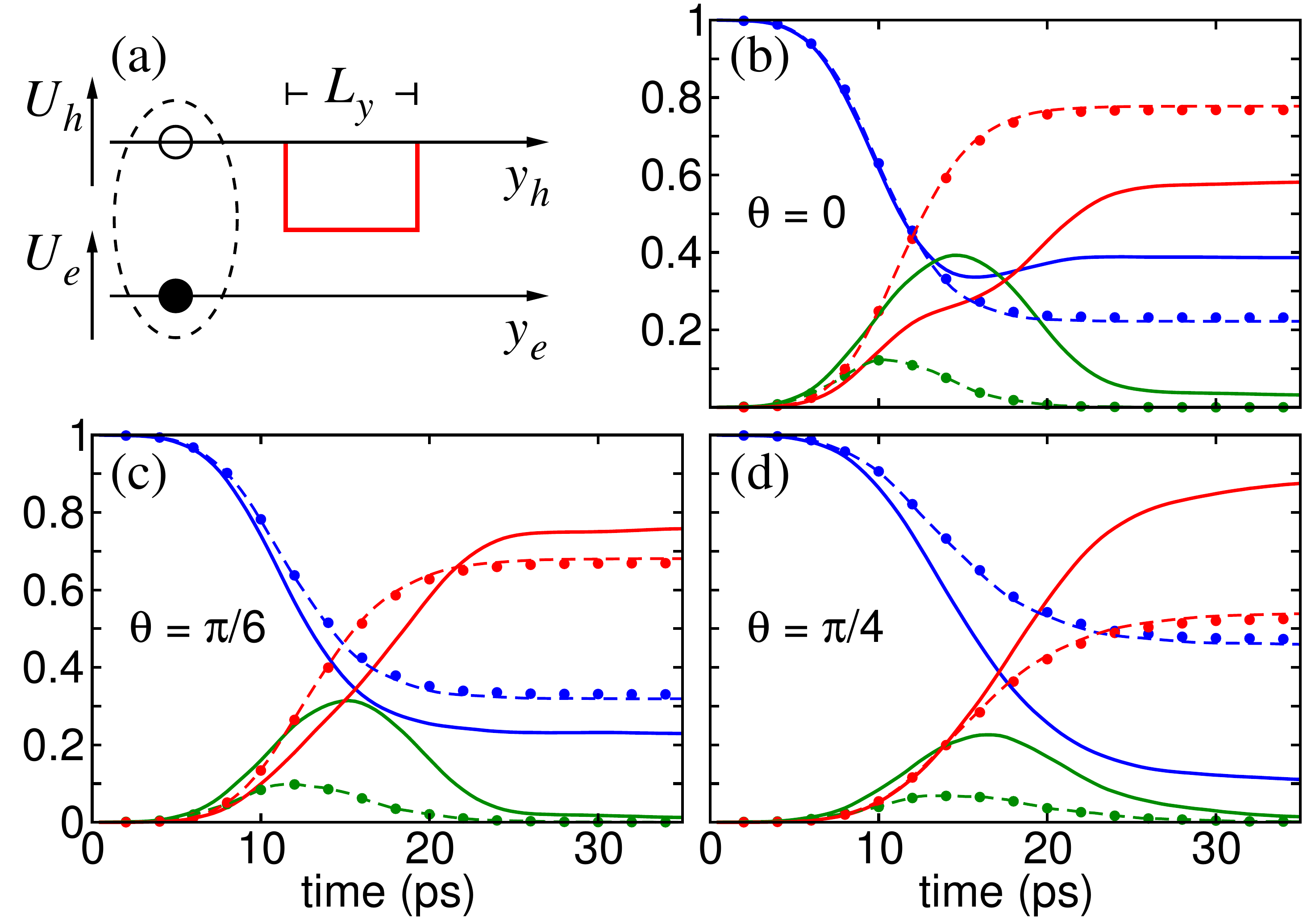}
\caption{Same as in Fig.~\ref{elec_well}, but with hole well $U_{h,0}=-3 \,\mbox{meV}$, $U_{e,0}=0$. Short animations of cases (b) and (d) are presented in the Supplemental Material.\cite{SupplMat_IX} }
\label{hole_well}
\end{figure}

We next consider the equivalent scattering problem, but with the potential well applied to the hole, with $U_{h,0} = -3.0 \,\mbox{meV}$, $U_{e,0}=0$.  Selected results are shown in Fig.~\ref{hole_well}(b-d). The kinetic energy and scattering angles are identical as in Fig.~\ref{elec_well}(a). Due to the different effective masses of electrons and holes, however, this problem is not equivalent to the previous one. 

Several differences with respect to the electron case (see Fig.~\ref{elec_well}) can be recognized. First, the component in the potential well during the scattering is larger and, accordingly, scattering times are longer. 
This is expected, due to the heavier mass of the hole.
However, the localization in the potential well (\textbf{B}-coefficient) is largely underestimated by the mean-field calculations, contrary to the previous case. 
Second, when the normal component of the exciton momentum is changed, there is no definite trend between the full calculation and mean-field calculations: at normal incidence the transmission coefficient is lower for the full propagation with respect to the RIX and TDH approaches. At $\theta=\pi/6$, Fig.~\ref{hole_well}(c), the \textbf{C}-coefficient is lower in the full propagation only until $t\approx 20\,\mbox{ps}$, but the asymptotic value is larger for the full calculation than in the mean-field approximations. 
Finally, at $\theta=\pi/4$, Fig.~\ref{hole_well}(d), the full \textbf{C}-coefficient is always larger than the RIX value, both during scattering and at asymptotic times. This should be ascribed to the activation of a resonant transmission channel.
This coupling is able to excite the IX into a superposition of higher internal states and it breaks the separability into CM and relative motion parts.
This is why even the TDH method is not able to reproduce the full results, and merely mimics the RIX approximation.   

A peculiar behavior is shown in Fig.~\ref{hole_well}(b) which exhibits a plateau in the \textbf{C}-coefficient during the scattering within the full dynamics. This indicates a non uniform transmission of the IX wave function. In a semiclassical picture, the hole is trapped into the well while the electron is partially transmitted. Due to Coulomb interaction, however, the transmitted electron inverts its motion, the IX bounds again and the pair is finally transmitted. Therefore, the transmitted wave-packet splits into an advanced and a delayed IX. Clearly, this requires the excitation of the internal modes, and this behavior is not reproduced by the RIX model.\cite{GrasselliJCP15} To confirm this interpretation, we calculate the classical internal oscillation period, (see Suppl. Mat.\cite{SupplMat_IX}) and estimate it in $\tau \approx 5.5 \, \mathrm{ps}$, which is indeed comparable to the time duration of the transmission plateau in Fig.~\ref{hole_well}(b).

\subsubsection{Symmetric potential barrier/well}

\begin{figure}
\includegraphics[width=\columnwidth]{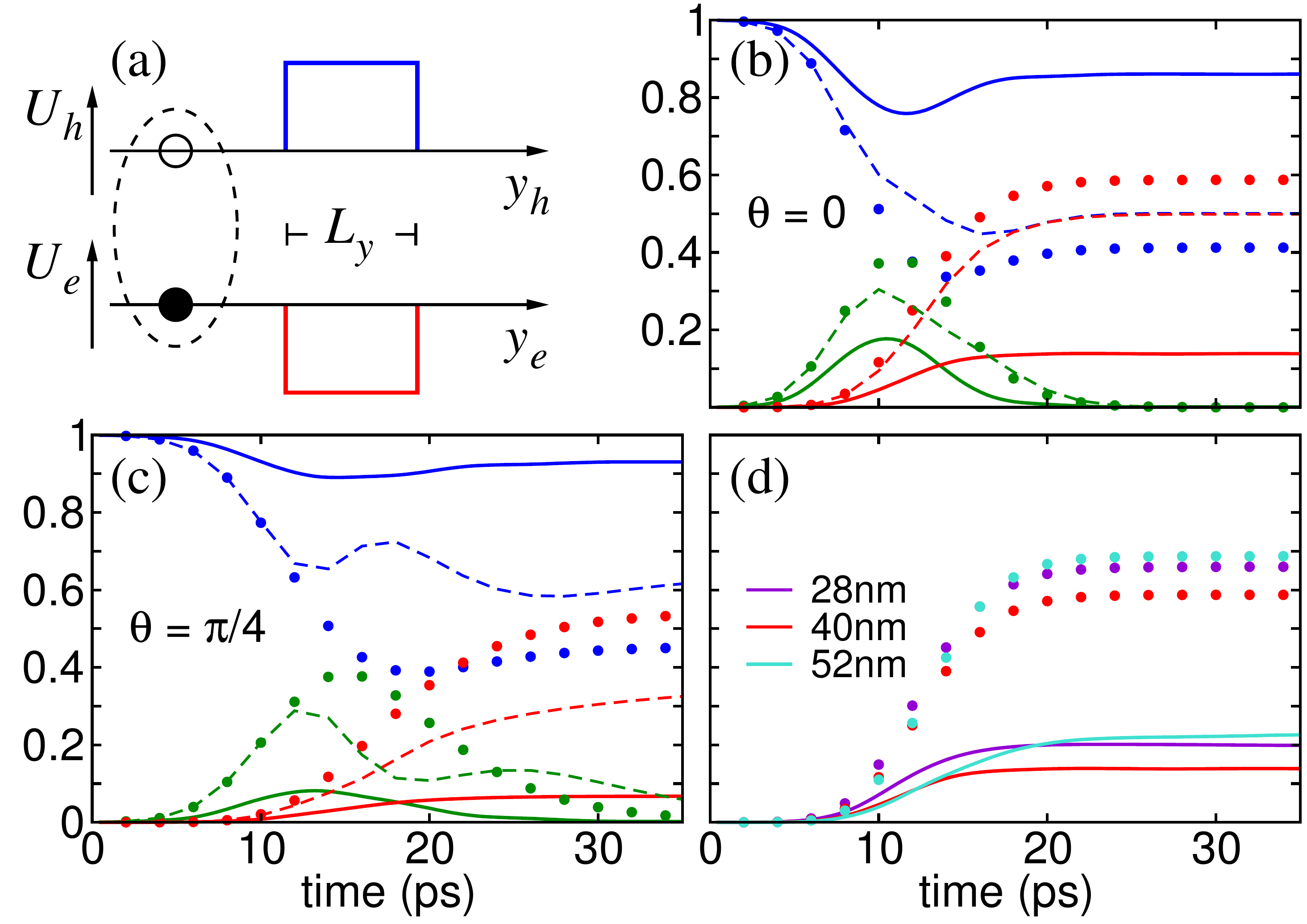}
\caption{Scattering in a symmetric potential $U_{h,0}=+3 \,\mbox{meV}$, $U_{e,0}=-3\,\mbox{meV}$. (a) Sketch of external potential. (b,c) Time evolution of the \textbf{A},\textbf{B},\textbf{C}-coefficients (line colors as in Fig.~3) at normal incidence $\theta=0$ and $\theta=\pi/6$, as indicated. Here, $L_y = 40 \,\mbox{nm}$. A short animation of case (b) is presented in the Supplemental Material\cite{SupplMat_IX} (d) Time evolution of the \textbf{C}-coefficient for potentials of different width $L_y$, as indicated. Dots: RIX approximation. Solid line: full propagation. In this case, the TDH approximation is not shown for clarity.} \label{gating_width}
\end{figure}

For completeness, we next discuss a symmetric potential with opposite sign for the two particles, $U_{h,0} = 3.0\,\mbox{meV}$ and $U_{e,0} = -3.0 \,\mbox{meV}$. The potential profile is sketched in Fig.~\ref{gating_width}(a). This is a somehow special situation, because the average potential is zero.

We consider $\theta=0$ and $\theta=\pi/4$ at the CM kinetic energy of $E_{\mbox{\scriptsize CM}}=0.4\,\mbox{meV}$. For both incident angles, the transmission coefficient for the full propagation is small, due to the repulsive barrier felt by the hole. Most interestingly, it is substantially smaller than the one obtained from the mean-field methods, for which the average of the external potential over the internal DoFs makes the hole barrier smooth, thus favoring transmission. 
Note also that for $\theta=\pi/4$ the TDH result deviates substantially from the RIX calculation. Nevertheless, it is still far from the full calculation.

To show that this is not an accidental situation, we have calculated the transmission coefficient as a function of the width of the external potential (Fig.~\ref{gating_width}(d)). Note that the behavior of the transmission coefficient is not monotonous with the the width of the external potential, $L_y$, as a possible consequence of resonant transmission for the CM DoFs alone. In particular, both for the full and the mean-field propagation, the transmission coefficient is larger for $L_y=28\,\mbox{nm}$ and $L_y=52\,\mbox{nm}$ than for $L_y=40\,\mbox{nm}$. Still, the transmission is systematically overestimated by the mean-field approach.

\subsection{Single slit potentials}
\label{sec:SingleSlit}

\begin{figure*}
\includegraphics[trim={0 0 0 8cm}, clip, width=\textwidth]{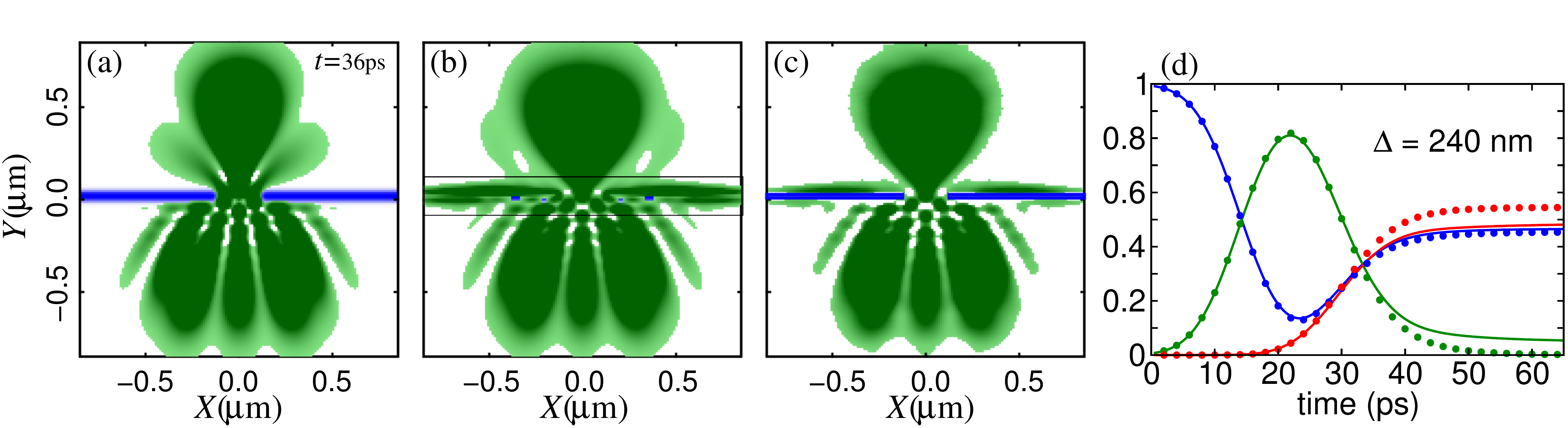} 
\caption{Scattering from a slit potential with $U_e = +6.0 \, \mbox{meV}, U_h = -3.0 \, \mbox{meV}, \, \Delta=240\,\mbox{nm}\, L_y=20\,\mbox{nm}$, $E_{\mbox{\scriptsize CM}}=0.4\,\mbox{meV}$ and normal incidence. A short animation of this case is presented in the Supplemental Material.\cite{SupplMat_IX} Snapshots of (a) the CM wave packet in the RIX approximation, (b) the CM marginal probability, and (c) $|\Psi(\mathbf{R},\mathbf{r}=\mathbf{0})|^2$ during scattering at $t=36\,\mbox{ps}$ are shown. Horizontal lines in panel (b) limit the integration domain for the potential region (see text). (d) Time evolution of the \textbf{A},\textbf{B},\textbf{C}-coefficients. Line colors as in Fig.~3. Dots: RIX approximation. Solid line: full propagation.}\label{fig8}
\end{figure*}

We finally investigate evolution through a single slit potential. This consists of an aperture of width $\Delta$ in an otherwise infinitely long barrier/well potential similar to those investigated in the previous sections. In Figs.~\ref{fig8} we summarize results for a slit with $\Delta=240\,\mbox{nm}$ and a slightly asymmetric electron/hole potential $U_e = 6\,\mbox{meV}$, $U_h = -3\,\mbox{meV}$, and $L_y = 20\,\mbox{nm}$. These values exclude tunneling through the barrier, while satisfying the condition $ U_{\mbox{\scriptsize ext}} \Delta_t/(2\pi\hbar) \ll 1$. The IX is initialized with a CM kinetic energy $E_{\mbox{\scriptsize CM}}=0.4\,\mbox{meV}$, moving towards the mid-point of the slit with normal incidence, and an initial width of the CM minimum uncertainty wave 
packet $\sigma_{X,Y}=160\,\mbox{nm}$.
These parameters produce several diffraction lobes in the transmitted wave-packet in the RIX approximation, shown in \ref{fig8}(a). Moreover they avoid the IX wave from spreading too much before reaching the slit.\cite{note_sigma} 

Figure~\ref{fig8}(b) shows a snapshot of the CM marginal probability of Eq.~(\ref{eq:MarginalProbability}) at $t=36\,\mbox{ps}$. A comparison with the equivalent RIX calculation in panel (a) helps to identify several important features. First, while the reflected part of the CM wave-packet is very similar in the two calculations, the diffraction lobes are almost suppressed in the full calculation. Second and most interesting, part of the CM wave-packet propagates as edge states along the barrier, far from the aperture. In semiclassical terms, this corresponds to a IX, with the hole trapped inside the well and the electron trapped on either side of the barrier by the electron-hole attraction. Therefore, this is a genuine correlation effect which cannot be reproduced by a mean-field approach, and indeed it is completely absent in Fig.~7(a). 
 
The evolution of the \textbf{A},\textbf{B},\textbf{C}-coefficients are shown in Fig.~\ref{fig8}(d) for full propagation and the RIX model (the TDH approximation almost coincides with the RIX one and it is not shown). Note that here the potential region for integration has been extended to ten times the external potential range ($200\,\mbox{nm}$, see Fig.~\ref{fig8}(a)) to measure accurately the asymptotic coefficients, since part of the wave function remains trapped at the edges of the potential, as we discussed above. These coefficients show a qualitative agreement between full and mean-field methods. This is because in the region where the wave function is large (the slit) the potential is vanishing. Therefore, the average external potential contribution is weak, and we are in a regime similar to Sec.~\ref{sec:WeakPotential}. However, at asymptotic times, the full calculation shows a transmission coefficient which is smaller than in the RIX calculation, the difference being the fraction of the wave function propagating along the edges of the external potential. Therefore, even if in a weak potential regime, correlation effects are exposed in the full calculation but not in the mean-field propagation. 

In Fig.~\ref{fig8}(c) we plot $ |\Psi(\mathbf{R},\mathbf{r}=\mathbf{0})|^2$ for the full calculation. On the one hand, there is very little difference with respect to the CM marginal probability of panel (b), indicating that the $\mathbf{r}=\mathbf{0}$ contribution is by far the largest in the relative coordinate average which provides the CM marginal probability. On the other hand, this is proportional to the IX optical recombination probability and shows that an optical luminescence experiment with sub-$\mu$m resolution would be able to probe the wave function with accuracy.\cite{MauritzPRL99}

We have repeated similar calculations, but with the electron and the hole potential exchanged, $U_e = -3.0 \, \mbox{meV}, U_h = +6.0 \, \mbox{meV}$. We observed, in this case, no qualitative difference with respect to the previous one. While full and mean-field calculations agree overall, there is a substantial part of the CM wave function which propagates as a bound IX along the potential edges, which is not captured by the mean-field calculation.

\begin{figure}
\includegraphics[clip, width=\columnwidth]{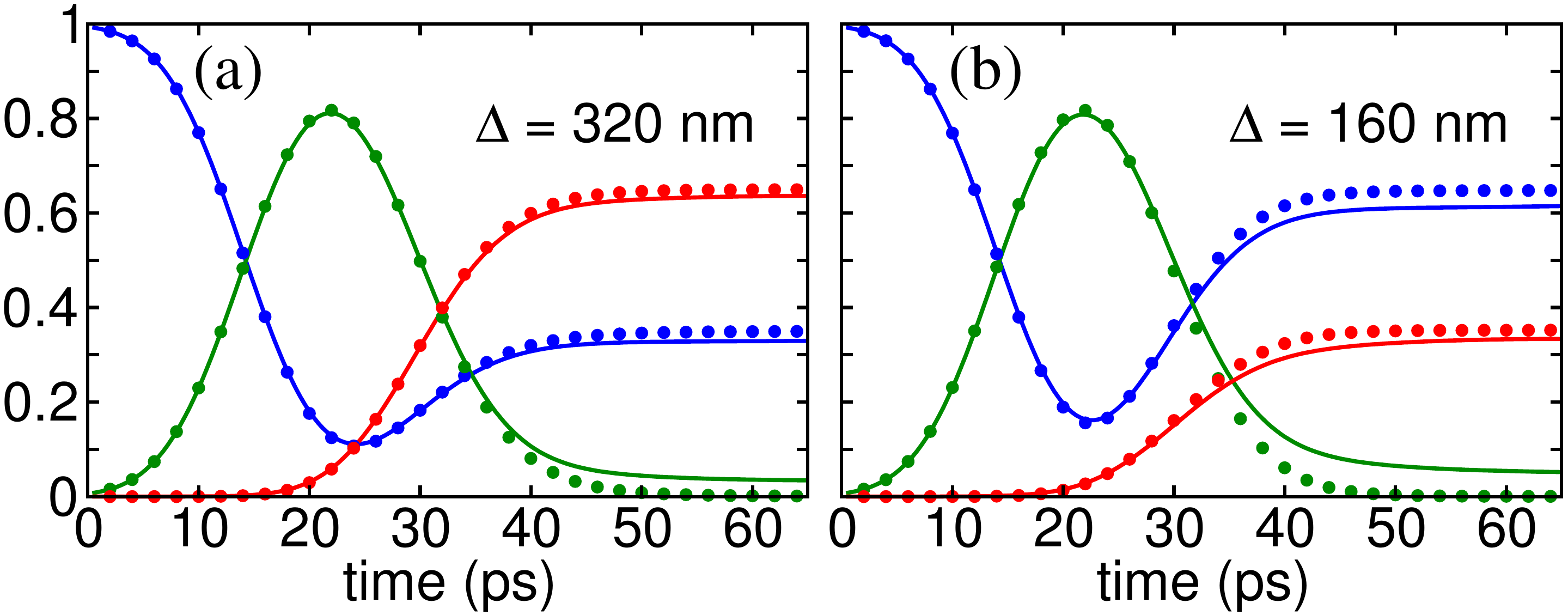}
\caption{Time evolution of the \textbf{A},\textbf{B},\textbf{C}-coefficients at selected values of $\Delta$, as indicated. Same potential strengths as in Fig.~7. Lines as in Fig.~7(d). Short animations of the propagation as in cases (a) and (b) are presented in the Supplemental Material.\cite{SupplMat_IX} }\label{fig9}
\end{figure}

In Fig.~\ref{fig8} we show a somehow special situation, where transmission and reflection probabilities are almost equal. Therefore, we show for completeness in Fig.~\ref{fig9} the evolution of the 
\textbf{A},\textbf{B},\textbf{C}-coefficients when the slit potential is `open', i.e., most part of the CM wave-packet is transmitted, or `closed', i.e., the CM wave-packet is almost fully reflected. Results are shown for the same potentials used in Fig.~\ref{fig8}. Again, we repeated the calculation interchanging the particle potential, finding no quantitative difference with the original one. Interestingly, in all cases a similar fraction of the CM wave function propagates along the edges of the potential barrier and well.
This phenomenon seems thus to be mostly related to the Bohr radius of the exciton, rather than the slit aperture: a larger (smaller) $\Delta$ determines the asymptotic value of the \textbf{B}-coefficient, at a given $\sigma$, only through the fact that a greater (smaller) part of the IX wave packet shall hit the edges of the slit.

\section{Conclusions}
\label{sec:Conclusions}

Scattering of composite particles is an important issue in several fields beyond semiconductors, such as molecular and nuclear scattering.\cite{Bertulani2015,AhsanPRC10} However, theoretical methods are relatively little developed, due to the numerical complexity. Here we have developed a numerical scheme to approach the particularly hard problem of a Coulomb bound complex scattering against arbitrary potentials, and we have applied the method to the specific case of IXs confined in CQW systems. 
On the one hand, this system offers the unique possibility to probe scattering of a composite particle with detailed optical means. On the other hand, we have shown here that \textit{full} numerical propagation can be obtained for this system, due to the limited number of DoFs allowed by the low dimensionality ensuing from CQW quantum confinement. This allows both to compare with perspective experiments and to test approximate and numerically simpler methods, such as mean-field methods.

For the present case we have shown that mean-field methods are predictive only for external potentials which, while coupling the CM and relative dynamics, are sufficiently weak, i.e., with an energy scale which is much smaller than the first excitation gap for the internal motion of the IX. In this case, the CM and relative DoF can be factorized at any time of the evolution.

The RIX and the TDH approximations, however, are inadequate in predicting the transmission and reflection coefficients for stronger potentials, i.e., when the strength of the external potential is comparable to internal levels spacing. In this situation the scattering potential partially excites (locally) the IX to higher internal levels, and the wave function cannot be factorized during evolution in the potential region. In such a case, the asymptotic IX transmission, as computed from the full propagation, can be larger or smaller than the transmission extracted from mean-field methods, depending on the specific parameters of the system. 

We have identified other signatures of the internal dynamics of the IX, coming into play during scattering, which cannot be captured by any mean-field calculation. For example, when plotted against time, the transmission through a well may exhibit plateaux before scattering is completed, which are absent in the simulations performed within RIX and TDH approximations. Moreover, in the single slit scattering problem, part of the wave-packet is not transmitted or reflected, but propagates along the edges. Again, this requires higher internal level excitations in the potential region through coupling of relative and CM DoFs which cannot be captured by a mean-field approach. This genuine correlation effect could be detected in optical experiments. It is also interesting to note that the RIX and TDH approximations basically coincide in almost all numerical simulations, indicating that, besides keeping the CM and relative subsystems decoupled, a mean field approach also smoothens the external potentials in such a way that it cannot exchange energy with the internal motion DoFs - which is \textit{de facto} always in its ground state - but only with the CM one.  

We finally note that, even for the case of only four DoFs, full wave-packet quantum propagation is a demanding task which required the development of a massively parallel code. On the one hand, exact calculations as the present one may serve as a severe benchmark for approximate, beyond mean-field methods which might be less computationally intensive. On the other hand, the implemented FSS method is suitable to treat also time-dependent external potentials, which might be produced in heterostructures for driving single carriers or bound IX wave packets, such as surface acoustic waves produced by interdigital devices, at small additional numerical cost.\cite{ViolanteNJP14} This is left as a future development.

\section{Acknowledgments}

We acknowledge INDEX for partial financial support. We acknowledge CINECA for computing time on parallel architectures under the Iscra C project IsC33-``FUQUDIX''. We acknowledge L. Butov and M. Fogler for useful discussions and suggestions during a stay at the University of California in San Diego, where part of this work has been developed.

%


\newpage
\setcounter{section}{0}
\onecolumngrid

\begin{center}

{\large {\bfseries Supplemental Material to ``Exact two-body quantum dynamics of an electron-hole pair in semiconductor coupled quantum wells: a time-dependent approach''}}\vspace*{\baselineskip}\\

\underline{Federico Grasselli},$^{1,2}$ Andrea Bertoni,$^2$ and
Guido Goldoni $^{1,2}$

\textit{$^1$Dipartimento di Scienze Fisiche, Informatiche e Matematiche, \\Universit\`a degli Studi di  Modena e Reggio Emilia, Via Campi 213/a, Modena, Italy}\\
\textit{$^2$S3, CNR-Istituto Nanoscienze, Via Campi 213/a, Modena, Italy}

\end{center}
\vspace{0.3cm}
\date{\today}

\section{The time dependent Hartree method}\label{App_TDH}

Consider a system composed of interacting subsystems 1 and 2, with the Hamiltonian\cite{note_App_1}
\begin{equation}
H(1,2) = H_1^{(0)}(1) + H_2^{(0)}(2) + U(1,2)
\end{equation}
where
\begin{equation}
H_j^{(0)}(j) = T_j(j) + U_j(j) \qquad j=1,2 
\end{equation}
are single subsystem Hamiltonians, and $T_j(j)$ and $U_j(j)$ are the kinetic energy and potential energies, respectively, for the $j-th$ subsystem. $U(1,2)$ is some potential which couples the coordinates for the two subsystems. 

The time-dependent Schr\"odinger equation reads
\begin{equation}
\Psi(1,2; t + dt) =\left(1 + \frac{dt}{i\hbar} H(1,2) \right) \Psi(1,2; t)
\end{equation} 
where $dt\rightarrow 0$.
We then \textit{assume} the following \textit{ansatz} for the total wave function
\begin{equation}
\Psi(1,2; t) = \eta(t)A(1;t)B(2;t), 
\qquad \forall t \in I\subseteq \mathbb{R}^+_0 ,
\end{equation}
i.e. a permanent separation during the time interval $I$ of the evolution.
Here we introduced the (redundant) variable $\eta=\eta(t)$ in order to be able of freely choose the global phases of $A(1,t)$ and $B(2,t)$. A constraint equation will thus be needed. 
By multiplying on the left by $\eta^*(t+dt)B^*(2;t+dt)$ and integrating on the variable set 2, we obtain
\begin{equation}
\begin{split}
A(1;t+dt) &= \int d2 \eta^*(t+dt)B^*(2;t+dt) \left(1 + \frac{dt}{i\hbar} H(1,2) \right) \eta(t)B(2;t) \;
A(1;t) \label{a1}
\end{split}
\end{equation}
In a similar way, we obtain
\begin{equation}
\begin{split}
B(2;t+dt) &= \int d1 \eta^*(t+dt)A^*(1;t+dt) \left(1 + \frac{dt}{i\hbar} H(1,2) \right) \eta(t)A(1;t) \;
B(2;t). \label{a2}
\end{split}
\end{equation}
and
\begin{equation}
\begin{split}
\eta(t+dt) &= \int d1 B^*(2,t+dt)A^*(1;t+dt) \left(1 + \frac{dt}{i\hbar} H(1,2) \right) A(1;t)B(2,t) \;
\eta(t). \label{a3}
\end{split}
\end{equation}
Consider Eq.~(\ref{a3}). We choose the phases of $A(1,t)$ and $B(2,t)$ such that the scalar products of these functions with their time derivatives, computed at the same time, vanish. This is equivalent, as $dt\rightarrow 0$, to 
\begin{equation}
\langle A(1;t+dt) | A(1;t) \rangle = \langle B(1;t+dt) | B(1;t) \rangle = 1, 
\end{equation}
$\forall t \in I$.
By inserting these equations into Eq.~(\ref{a3}), we have the following equation of motion for $\eta(t)$:
\begin{equation}
\begin{split}
\eta(t+dt) &= \int d1 B^*(2,t)A^*(1;t) \left(1 + \frac{dt}{i\hbar} H(1,2) \right) A(1;t)B(2,t) \;\eta(t) = \\
& = \left(1+\frac{dt}{i\hbar}\langle 1,2| H(1,2) |1,2\rangle (t)]\right) \eta(t)
\end{split}
\end{equation}
i.e. $i\hbar\partial_t\eta =  \langle H \rangle \eta$, which can be itself seen as \textit{the} constraint equation, and has as a (formal) solution
\begin{equation}
\eta(t) = \mathcal{T} \exp \left( \frac{1}{i\hbar} \int_0^t [\langle 1,2| H(1,2) |1,2\rangle (t')] dt' \right)
\end{equation}
This, by the way, is just a global-in-space phase and thus it is of no importance when we're interested in the time evolution of the \textit{probability density}.
With this constraint, we then have the following expression for Eq.~(\ref{a1})
\begin{equation}
\begin{split}
A(1;t+dt) &= \left\lbrace 1 + \frac{dt}{i\hbar} \left( H_1^{(0)}(1) + \beta(t) +  u_{\mathrm{1},\mbox{\scriptsize eff}}(1;t) - [\langle 1,2| H(1,2) |1,2\rangle (t)] \right) \right\rbrace A(1;t) 
\end{split}
\end{equation}
where
\begin{equation}
\beta(t) \equiv \int d2 B^*(2;t)H_2^{(0)}(2)B(2;t),
\end{equation}
and
\begin{equation}
u_{\mathrm{1},\mbox{\scriptsize eff}}(1;t) \equiv \int d2 |B(2,t)|^2 U(1,2).
\end{equation}
The last term, $[\langle 1,2| H(1,2) |1,2\rangle (t)]$, comes from the constraint equation.
We must specify that, if one is interested in probability densities, the global-in-space `effective potentials' $\beta(t)$, $[\langle 1,2| H(1,2) |1,2\rangle (t)]$ are not influential,\cite{note_App_2} and thus we can re-write the equation of motion for the variable $A(1,t)$ simply as
\begin{equation}
A(1;t+dt) = \left[ 1 + \frac{dt}{i\hbar} \left( H_1^{(0)}(1) +  u_{\mbox{\scriptsize eff}}(1;t) \right) \right] A(1;t) 
\end{equation}
that is
\begin{equation}
i\hbar \partial_t A(1;t) = \left( H_1^{(0)}(1) +  u_{\mbox{\scriptsize eff}}(1;t) \right) A(1,t)
\end{equation}
Analogously, for the variable $B(2;t)$ we find
\begin{equation}
i\hbar \partial_t B(2;t) = \left( H_2^{(0)}(2) +  u_{\mathrm{2},\mbox{\scriptsize eff}}(2;t) \right) B(2,t)
\end{equation}
where
\begin{equation}
u_{\mathrm{2},{\mbox{\scriptsize eff}}}(2;t) \equiv \int d1 |A(1;t)|^2 U(1,2).
\end{equation}
We can formally solve the latter equations by means of the evolution operator:
\begin{align}
A(1,t+\Delta_t) = \mathcal{U}_1(1; t+\Delta_t, t) A(1;t) , \label{eqTDH1}\\
B(2,t+\Delta_t) = \mathcal{U}_2(2; t+\Delta_t, t) B(2;t) \label{eqTDH2}
\end{align}
where
\begin{equation}
\mathcal{U}_1(1; t+\Delta_t, t) = \exp \left[ \frac{1}{i\hbar} \int_t^{t+\Delta_t} dt' \left( H_1^{(0)}(1) +  u_{\mbox{\scriptsize eff}}(1;t')  \right)  \right].
\end{equation}
and
\begin{equation}
\mathcal{U}_2(2; t+\Delta_t, t) = \exp \left[ \frac{1}{i\hbar} \int_t^{t+\Delta_t} dt' \left( H_2^{(0)}(2) +  u_{\mbox{\scriptsize eff}}(2;t')  \right) \right].
\end{equation} 
which can be handled by the split-step-Fourier method.

To summarize, the quantum evolution of a system composed of subsystems 1 and 2 have been separated in the quantum evolutions each subsystem separately, each with a time-dependent Hamiltonians. For, say, subsystem 1, an effective potential $u_{\mbox{\scriptsize eff}}(1;t)$ arises, which can be written as the expectation value of the coupling potential averaged on the wave function of subsystem 2. Equivalently for particle 2, with particle coordinates exchanged. In this sense, this is a mean-field-like approach. 

The Fourier split step method (FSS) for each of the two subsystems can then be applied to obtain the global evolution. In the present case, the systems 1 and 2 are representing the CM and relative motion coordinates, respectively; in this case, Eqs.~(\ref{eqTDH1}) and (\ref{eqTDH2}) read as Eqs.~(21) and (22) in the main text:
\begin{equation}
\begin{split}
\chi(\mathbf{R};t+\Delta_t) &= \exp \left\lbrace - \frac{i}{\hbar} \int_t^{t+\Delta_t}dt'  \left[ \frac{\mathbf{P}^2}{2M} +  U_{\mbox{\scriptsize eff}}(\mathbf{R};t')   \right] \right\rbrace \chi(\mathbf{R};t) 
\end{split} \label{eqTDHCM}
\end{equation}
and
\begin{equation}
\begin{split}
\phi(\mathbf{r};t+\Delta_t) &= \exp \left\lbrace -\frac{i}{\hbar} \int_t^{t+\Delta_t} dt'  \left[ \frac{\mathbf{p}^2}{2m} + U_{C}(\mathbf{r}) + u_{\mbox{\scriptsize eff}}(\mathbf{r};t') \right] \right\rbrace \phi(\mathbf{r};t) 
\end{split} \label{eqTDHrel}
\end{equation}
with
\begin{eqnarray}
U_{\mbox{\scriptsize eff}}(\mathbf{R};t) \equiv \int d\mathbf{r} |\phi(\mathbf{r};t)|^2 U_{\mbox{\scriptsize ext}}(\mathbf{R},\mathbf{r})  
\label{effTHDpotss}
\\
u_{\mbox{\scriptsize eff}}(\mathbf{r};t) \equiv \int d\mathbf{R} |\chi(\mathbf{R};t)|^2 U_{\mbox{\scriptsize ext}}(\mathbf{R},\mathbf{r}) \label{effTHDpots}
\end{eqnarray}

\section{Classical internal oscillation period for the first excited ($p$) state}

Starting from the expression of the classical energy for the first excited state (we choose for example the $p_y$ state, whose eigen-energy is $\mathcal{E}_1$) 
\begin{equation}
\mathcal{E}_1 = \frac{1}{2} m \dot{y}^2 + U_C(y) \label{eqEnClassCons}
\end{equation}
(we assume for simplicity the trajectory at fixed $x=0$), where $U_C(y) \equiv e^2/(4\pi\epsilon_0\epsilon_r\sqrt{d^2+y^2})$, and taking $d$ as the distance between the centers of the two wells along the growth axis, we have the classical turning points (at which $\dot{y}=0$) $y = \pm \overline{y}$, where $\overline{y}\equiv \sqrt{[e^2/(4\pi\epsilon_0\epsilon_r\mathcal{E}_1)]^2 - d^2} $.
We can then compute the classical internal oscillation period, $\tau$, by isolating $(\dot{y}^2)^{-1/2}$ in Eq.~(\ref{eqEnClassCons}), integrating over $y$ between 0 and $\overline{y}$, and multiply the result by four, since the integration corresponds to one fourth of the classical path \cite{LandauMech}: 
\begin{eqnarray}
\tau = 4 \int_0^{\overline{y}} dy \left[\frac{2}{m} \left(\mathcal{E}_1 + \frac{e^2}{4\pi\epsilon_0\epsilon_r\sqrt{d^2+y^2}}\right)\right]^{-1/2}, 
\end{eqnarray}
Taking the specific values concerning the physical system adopted in the manuscript, we have $d=12\,\mbox{nm}$, and $\mathcal{E}_1=-1.28\,\mbox{meV}$ (see Tab.~II in the main text), and the (numerical) integration leads to $\tau \approx 5.5 \, \mathrm{ps}$.


\begin{thebibliography}{52}
\bibitem []{note_gating}%
  \BibitemOpen
  \href@noop {} {} \bibinfo {note} {It is in principle
  possible to gate the two layers independently with top and bottom gates,
  generating potentials of the same sign for the two
  particles. However, these complex
  configurations will not be investigated in this paper. See also Ref.
  \protect\onlinecite{GrasselliJCP15}.}\BibitemShut {Stop}%
\bibitem []{note_sigma}%
  \BibitemOpen
  \href@noop {} {} \bibinfo {note} {Since
  $Y_0=-600\,\mbox{nm}$, at $t^*\equiv (|Y_0| +
  L_y/2)/\sqrt{2E_{\mbox{\scriptsize CM}}/M} \approx 21.7\,\mbox{ps}$, which is
  the time needed by the center of the CM wave-packet to reach the mid-point of
  the slit, $\sigma_{X,Y}(t^*) =
  \sigma_{X,Y}(t=0)\sqrt{1+\{t^*\hbar/[2M\sigma^2_{X,Y}(t=0)]\}^2} \approx
  166\,\mbox{nm}$, which is very similar to the intial value
  $160\,\mbox{nm}$.}\BibitemShut {Stop}%
\end{thebibliography}
\begin{thebibliography}{52}
\bibitem [2]{note_App_2}%
  \BibitemOpen
  \href@noop {} {} \bibinfo {note} {It is nothing but a
  global gauge transformation.}\BibitemShut {Stop}%
\end{thebibliography}
\end{document}